\documentstyle{article}
\begin{document}
\title{${\bf{Z}}_2$-Graded Cocycles in Higher Dimensions}
\author{{\bf C. Ekstrand}\\Department of Theoretical Physics, \\Royal Institute of
Technology, \\S-100 44 Stockholm, Sweden}
\date{}
\maketitle

\newcommand{\eq}{\begin{equation}}
\newcommand{\eqend}{\end{equation}}
\newcommand{\eqa}{\begin{eqnarray}}
\newcommand{\eqaend}{\end{eqnarray}}
\newcommand{\nonu}{\nonumber \\ \nopagebreak}
\newcommand{\Ref}[1]{(\ref{#1})}

\newcommand{\B}{{\cal B}}
\newcommand{\F}{{\cal F}}
\newcommand{\cL}{{\cal L}}

\begin{abstract}
Current superalgebras and corresponding Schwinger terms in 
$1$ and $3$ space dimensions are studied. This is done by generalizing the 
quantization of 
chiral fermions in an external Yang-Mills potential to the case of a 
${\bf{Z}}_2$-graded potential coupled to bosons and fermions.
\end{abstract}
\section{Introduction}
Since the suggestion of supersymmetry 
 as a possible symmetry 
in nature there has been an increased interest in theories allowing for a 
${\bf{Z}}_2$-graded extension.
 In this paper we will find ${\bf{Z}}_2$-graded generalizations of the 
Lundberg\cite{LU} and Mickelsson-Faddeev\cite{M,F}
cocycle, and the underlying formalism.
 The cocycles under consideration appear in the representation theory of the 
infinite dimensional Lie algebras  $\mbox{Map}(M;{\bf{g}}) $ of smooth maps from a 
$ d$-dimensional compact manifold $M$ to a 
compact, semi-simple Lie algebra ${\bf{g}}$ (e.g. ${\bf{g}}=su(N)$). These
 maps can be interpretated as gauge transformations in some gauge theory.  

 When $M$ is a $C^\infty$ manifold with a Riemannian- and a spin structure 
there is a natural embedding of $\mbox{Map}(M;{\bf{g}}) $ and ${\cal A}(M)$, 
the set of Yang-Mills configurations, into some sets ${\bf{g}}_p$ and 
 $Gr_p$, $p=(d+1)/2$, respectively. How this is made will be described later. 
 The Lie algebras ${\bf{g}}_p$ and Grassmannians $Gr_p$ 
are used in the algebraic investigation of the representations of 
$\mbox{Map}(M;{\bf{g}})$.
To define them one considers a separable Hilbert space $h$ and a grading 
operator $F_0$ (i.e. $F_0=F_0^{\ast}=F_0^{-1}$; the star denotes the Hilbert 
space adjoint) such that the spaces $h_{\pm}=
(1\pm F_0)h$ are infinite dimensional. Then  ${\bf{g}}_p$ is defined as the 
Lie algebra of all bounded operators $X$ on $h$ such that 
$([F_0,X]^{\ast}[F_0,X])^p$ is trace class (we recall, 
see ref.~\cite{S}, that an 
operator $a$ on $h$ is trace class if $\sum _n|(f_n,af_n)|$ 
is finite for an arbitrary orthonormal basis $\{ f_n\}$ in $h$, and in 
that case its Hilbert space trace $tr(a)=\sum _n(f_n,af_n)$ exists, i.e. it is 
finite and basis independent). Similarly, $Gr_p$ is defined as 
the set of grading operators $F$ such that $((F-F_0)^{\ast}(F-F_0))^p$ 
is trace class.
 The embedding described above is then possible for the case $h=L^2(M;V)$, 
 where $V$ is a representation space of ${\bf{g}}$.

 The representation theory for the case $p=1$ ($d=1$) is 
well-understood (see for example ref.~\cite{I}).
The reason is that elements in 
${\bf{g}}_p$ can only be implemented (or second quantized) in the physical 
relevant representation of the field algebra if $p = 1$.
 It is also known how to make a ${\bf{Z}}_2$ extension of the $p=1$ 
 case and a graded Lundberg cocycle has been obtained\cite{GL}.

In higher dimensions the situation is much more difficult. 
A straight forward attempt to second quantize elements in ${\bf{g}}_p$, $p>1$ 
will fail. One will have to deal with infinities corresponding to the 
divergencies arising in certain Feynman diagrams in gauge theory models.
 There exists several different 
renormalization methods to handle such divergencies. The one we will use is 
close to the one used in  ref.~\cite{L}.
 Here the implementations is by forms obtained by an appropriate 
multiplicative regularization. For this, the Grassmannians $Gr_p$ needs to 
be introduced. The case $d=3$ and chiral fermions leads to the 
Mickelsson-Rajeev cocycle\cite{MR}. A corresponding boson cocycle has 
been obtained in ref.~\cite{La}.

In this paper we will generalize the results described above to a 
${\bf{Z}}_2$-graded case. A local form of the cocycles will be 
calculated, and some applications motivated by gauge theory models will 
be considered. 
\section{Preliminaries}
To fix notation we summarize some basic facts about  
${\bf{Z}}_2$-graded vector spaces and algebras. An element $v$ in a  
${\bf{Z}}_2$-graded vector space $V=V_{\bar{0}}\oplus V_{\bar{1}}$ is said 
to be homogeneous of  degree  $\alpha $, $\mbox{\small{deg}}(v)
=\alpha$, if $v\in V_{\alpha }$,  $\alpha \in  {\bf{Z}}_2\equiv 
\{ \bar{0}, \bar{1} \} $. 
If $V$ is also an 
algebra with grading preserving multiplication, i.e. 
 $v\in V_{\alpha }, w\in V_{\beta } \Rightarrow vw\in V_{\alpha + \beta}$
, then it is called a  
${\bf{Z}}_2$-graded algebra.
We define the supercommutator 
$\left[ \cdot , \cdot \right] _s :$ $V\times V \rightarrow V$, to be the
bilinear map 
\eq
\label{eq:SC}
\left[ v , w\right] _s = v w -(-1)^{\mbox{\small{deg}}(v)
\mbox{\small{deg}}(w)}wv.
\eqend
 This formula, as well as all other formulas in the paper, is by linearity 
also defined for non-homogeneous elements.
 Equipped with the supercommutator, $V$ becomes a Lie superalgebra (this 
subject is described in ref.~\cite{K}).
Every linear operator $X$ on $V$ can be written 
in matrix form 
\eq
\label{eq:MX}
X=\left(
 \begin{array}{cc}
 X_{\bar{0}\bar{0}} &  X_{\bar{0}\bar{1}} \nonu
  X_{\bar{1}\bar{0}} &  X_{\bar{1}\bar{1}} 
 \end{array} \right)
\eqend
corresponding to the decomposition $V=V_{\bar{0}}\oplus V_{\bar{1}}$. 
Then  $\mbox{\small{deg}}( X_{\alpha\beta })=\alpha +\beta$ defines a 
grading which provides every algebra of linear operators on $V$ with a
${\bf{Z}}_2$-structure.

We will now consider a ${\bf{Z}}_2$-graded Hilbert space and define certain 
operators acting thereon.
 Let $h=h_{\bar{0}}\oplus h_{\bar{1}}$ be an infinite dimensional separable 
${\bf{Z}}_2$-graded Hilbert space with the subspaces $h_{\bar{0}}$ and 
 $h_{\bar{1}}$ both infinite dimensional.
 Let $P_{\alpha}$ denote the orthogonal 
projection onto $h_{\alpha }$: $h_{\alpha}=P_{\alpha}h$, and introduce the 
Klein operator $\gamma = P_{\bar{0}}-P_{\bar{1}}$. 
We denote by $\B$ the algebra of bounded operators on $h$ and by $\B _{2p}$ 
the Schatten ideal classes (see  ref.~\cite{S} consisting of operators $X\in\B$ such that 
$\left(X^\ast X\right)^p$ has a converging Hilbert space trace 
 (or supertrace, which will be defined later).
Especially, $\B _1$ and  
$\B _2$ are the ideal of trace class and Hilbert-Schmidt operators, 
respectively. 
We define
\eq
Gr_{p}\left( F_0 \right) = \{ F\in \B; F^2=1 \quad \mbox{and} \quad 
F-F_0\in\B_{2p}\}
\eqend
for some fixed operator $F_0\in \B$ obeying 
\eqa
[F_0,\gamma ] & = & 0 \nonu
F_0  =  F_0^{\ast } & = & F_0^{-1}.
\eqaend
 The operator $F_0$ will be thought as being the sign of the free Hamiltonian. 
 It is also useful to introduce the classes
\eq
{\bf{g}}_p(F) = \{ X\in \B ;
\left[ F,X \right] _s
\in  \B _{2p} \} .
\eqend

 We regard the subspaces $h_{\bar{0}}$ and $h_{\bar{1}}$ as the 
one-particle spaces of charged bosons and fermions, respectively. Using the 
bosonic and fermionic Fock spaces $\F _B(h_{\bar{0}})$ and 
$\F _F(h_{\bar{1}})$ (defined in  ref.~\cite{I} for example) with vacua 
$\Omega_B$ and  $\Omega_F$ we 
construct the Fock space $\F _{\gamma} (h) = \F _B(h_{\bar{0}})\otimes 
\F _F(h_{\bar{1}})$ with vacuum $\Omega = \Omega_B\otimes\Omega_F$. 
 The bosonic and fermionic creation and annihilation operators  
can be combined to the operators
\eq
a(f)  =  a_B ( P_{\bar{0}}f)\otimes {\bf 1}_{\bar{1}} 
+{\bf 1}_{\bar{0}} \otimes a_F ( P_{\bar{1}}f)
\eqend 
and $a^{\dagger }(f)=a(f)^{\ast}$, $ \forall f\in h$, the Fock space adjoint, 
 which act on $\F _{\gamma} (h) $. 
They obey $a(f)\Omega =0$, and have a common, dense and invariant domain in 
$\F _{\gamma} (h)$. Introducing the grading 
\eq
\mbox{\small{deg}}\left( a^{\dagger }( P_{\bar{\alpha }}f) 
\right)= \alpha =\mbox{\small{deg}}\left( a( P_{\bar{\alpha }}f) 
\right), \quad
\alpha \in {\bf{Z}}_2, f\in h
\eqend
we can extend the canonical commutation relations (CCR) and anticommutation 
relations (CAR) of the bosonic and fermionic creation and 
annihilation operators to the canonical supercommutator 
relations\cite{GL} (CSR)
\eq
\label{eq:CSR}
\left[ a(f), a^{\dagger }(g)\right]_s =\left( f,g \right) {\bf 1} \quad 
\left[ a(f),a(g)\right]_s =0 \quad \forall f,g \in h
\eqend 
 where  $( \cdot ,\cdot )$ is the scalar product in $h$. 
 We now define the CSR algebra over $h$ to be the quotient of the free 
 $\ast$ algebra with complex coefficients, generated by $a(f)$, $a^{\dagger} 
(f)$, $f\in h$, and the identity ${\bf 1}$, by (the two sided $\ast$ algebra 
generated by) $a^{\dagger }(f)=a(f)^{\ast}$, $a^{\dagger }(c_1f_1+c_2f_2)=
c_1a^{\dagger }(f_1)+c_2a^{\dagger }(f_2)$ and the relations in \Ref{eq:CSR}. 
 A representation of this algebra was constructed above, usually referred to 
as the free (Fock-Cook) representation of the CSR algebra.
 From  a physical point of view this representation is not satisfactory 
since it provides us with a Hamiltonian operator that is unbounded from 
below.  We will instead consider a representation
where this is avoided (a so called highest weight representation -- 
see ref.~\cite{KR}).
 Using the elements in the free representation 
the so called quasi-free representation of
 the CSR algebra can be constructed:
\eqa
\label{eq:QF}
a(f;F_0) & = & a(P^{F_0}_+ f)+a^{\dagger}(JP^{F_0}_-f) \nonu
a^{\dagger}(f;F_0) & = & a^{\dagger}(P^{F_0}_+ f)-a(\gamma JP^{F_0}_-f)
\eqaend
where $J$ is a conjugation in $h$ (an antilinear norm-preserving operator 
obeying $J^2=1$) commuting with $\gamma$ and with 
$P^{F_0}_{\pm}=(1\pm F_0)/2$. 
 The procedure of going from the free to the quasi-free 
representation is a generalization of the well known process of 
\lq filling the Dirac sea\rq . 
\section{Second Quantization and ${\bf Z}_2$-graded cocycles}
For certain bounded operators $X$ on $h$ the corresponding Fock space 
operator $d\Gamma (X;F_0)$ can be defined. It is by definition the operator 
on $\F _\gamma(h)$ that has vanishing vacuum expectation value
\eq 
\label{eq:VE}
<\Omega , d\Gamma (X;F_0)\Omega >=0
\eqend
and satisfies
\eq
\label{eq:GA}
\left[ d\Gamma (X;F_0), a^\dagger (f;F_0) \right] 
_s  =  
a^\dagger (Xf;F_0) ,\quad f\in h.
\eqend
 where $<\cdot ,\cdot >$ denotes the scalar product in $\F _\gamma (h)$. 
The relation \Ref{eq:GA} defines $d\Gamma (X;F_0)$ up to an additive 
c-number possibly  depending on $X$.
 One way to fix the value of the c-number is by using eq. \Ref{eq:VE}. 
Let  $\{ f_n \} _{n=-\infty }^\infty $ be a complete system of orthonormal 
 homogeneous vectors in $h$ with 
$F_0f_n=\lambda _n f_n$, where the eigenvalues are indexed such that 
$\lambda _{n<0} =-1$ and  $\lambda _{n\geq 0} =1$. Then it is easy to see 
that if the operator  $d\Gamma (X;F_0)$ exists, it must be of form 
\eq
d\Gamma (X;F_0) = \quad \sum_{n,m=-\infty }^\infty (f_n, X f_m)
:a^{\dagger} (f_n;F_0) a (f_m;F_0):
\eqend
where the normal ordering is defined by 
 \eq
:a^{\dagger} (f_n;F_0) a (f_m;F_0):\quad =\left\{ \begin{array}{ll}
 (-1)^{\mbox{\small{deg}}(f_m) } a (f_m;F_0)a^{\dagger} (f_n;F_0) & \mbox{if}\quad n=m<0 \\
 a^{\dagger} (f_n;F_0) a (f_m;F_0) & \mbox{otherwise.}
 \end{array}\right.
\eqend
 The necessary and sufficient condition 
for existence of  $d\Gamma (X;F_0)$ is that $X\in {\bf{g}}_1(F_0)$\cite{GL}.

The operator $X$ will throughout the paper be thought of as the generator 
of an infinitesimal gauge transformation. The wave functions on which it acts 
are assumed to be dependent on some $F\in Gr_1(F_0)$ and second quantized 
according to the polarization in $h$ given by $F_0$. To take into account the 
action a gauge transformation has on $F$ (corresponding to its action on the 
Hamiltonian) we introduce the operator 
$G(X;F_0)=d\Gamma (X;F_0)+\cL _X$, with the Lie derivative $\cL _X, X\in 
 {\bf{g}}_1(F_0) $ acting on functionals  $m(F), F\in Gr_1(F_0)$, as 
\eq
(\cL _Xm)(F)= \frac{d}{dt}m\left( F-t\left[ F,X \right] _{(s)} 
+ {\cal O}(t^2) \right) |_{t=0}.
\eqend
 Taking the supercommutator of two $G$'s gives the relation
\eq
\label{eq:GG}
\left[ G(X;F_0), G(Y;F_0) \right]_s = G\left( \left[ X,Y\right] _s;F_0 
\right) - c(X,Y;F_0), \quad X,Y\in{\bf{g}}_1(F).
\eqend 
The Schwinger term 
\eq
\label{eq:ST}
c(X,Y;F_0)  =  d\Gamma \left(\left[ X,Y\right] _s;F_0\right)-
\left[ d\Gamma \left( X;F_0\right) 
d\Gamma \left( Y;F_0\right) \right] _s .
\eqend 
can be calculated by taking the vacuum expectation value of \Ref{eq:ST} 
and performing some algebraic manipulations. The result is 
\eqa
\label{eq:LC}
c_1(X,Y)=c(X,Y;F_0) & = & -\frac{1}{4}\mbox{str}\left( F_0
\left[ F_0,X\right] _{(s)}\left[ F_0,Y\right] _{(s)}\right)\nonu
& = &
-\frac{1}{2}\mbox{str}_{\mbox{c}}\left( X\left[ F_0 , Y\right] _{(s)}\right).
\eqaend 
 Since $\mbox{\small{deg}} (F_0) = \bar{0}$ our 
 notation $(s)$ is to illustrate that the commutator as well as 
the supercommutator may be used in the equations.
 The conditional trace,  conditional supertrace and 
ordinary supertrace can be defined in terms of the ordinary trace 
according to
\eqa
\mbox{str}_{\mbox{c}}(a) &= & \mbox{tr}_{\mbox{c}}(\gamma a)=
\frac{1}{2}\mbox{tr}\left(\gamma (a+F_0aF_0)\right) 
,\quad a+F_0aF_0\in\B _1 \nonu
\mbox{str}(a) &= & \mbox{tr}(\gamma a) \quad a\in\B _1.
\eqaend
The term $c$ is a graded generalization of the Lundberg cocycle\cite{LU}. 
The graded Jacobi identity is equivalent with the 2-cocycle relation 
\eq
\label{eq:CC1}
c([X,Y]_s,Z; F_0)+ \mbox{graded cyclic perm.} =0, \quad X,Y,Z\in{\bf{g}}_1
(F_0).
\eqend 
 If a different condition than \Ref{eq:VE} had been used to determine the 
$c$-number ambiguity in \Ref{eq:GA}, then $G(X;F_0)$ 
would have been changed to $G(X;F_0)+\zeta (X)$ for some complex valued 
function $\zeta$  of ${\bf{g}}_1(F_0)$. This would in turn change the 
cocycle $c$ to $c + \delta \zeta$, with
\eq
\label{eq:CB1}
\delta\zeta (X,Y)=\zeta ([X,Y]_s).
\eqend
The new cocycle is in the same cohomology class as the old one since 
they differ only by a function of type $\delta \zeta$, a coboundary. 
 
 When considering the corresponding situation for some $F\in Gr_1(F_0)$ 
we define $\tilde{G}(X,F), X\in {\bf{g}}_1(F)$ (or equivalently 
$ X\in {\bf{g}}_1(F_0)$) by
\eq
\tilde{G}(X,F)={\cal U}^{\ast}(F)G(X,F_0){\cal U}(F)
\eqend 
for some unitary operator ${\cal U}(F)$ on $\F _\gamma(h)$. Obviously it 
obeys a relation as the one in \Ref{eq:GG} and with the same Schwinger term. 
 Changing $\tilde{G}(X,F)$ by a $c$-number $\zeta (X;F)$ gives a new 
kind of coboundary: 
\eq
\label{eq:CB}
\delta\zeta (X,Y;F)=\zeta ([X,Y]_s;F)-\cL _X\zeta (Y;F)+
(-1)^{\mbox{\footnotesize{deg}}(X) \mbox{\footnotesize{deg}}(Y)}
\cL _Y\zeta (X;F).
\eqend
Since the Schwinger term then can be $F$-dependent, the 2-cocycle relation 
will now be
\eq
\label{eq:CC}
c([X,Y]_s,Z; F)-{\cal L}_Xc(Y,Z; F) 
+ \mbox{graded cyclic perm.} =0, \quad X,Y,Z\in{\bf{g}}_1(F).
\eqend

 $d\Gamma (X;F)$ does no longer exist as an operator when $X\in{\bf{g}}_2(F)$ 
and $F\in Gr_2(F_0)$. Therefore some additional regularization is needed. 
 A useful observation is that $d\Gamma (X;F)$ can still be 
defined as a sesquilinear form\cite{R1,R2}.
Proceeding as in the ungraded case\cite{L} we make use of this 
by finding a coboundary 
  $(\delta b) (X,Y;F)$ for  $X,Y\in{\bf{g}}_1(F)$, $F\in Gr_1(F_0)$ 
such that $(\delta b) (X,Y;F)$ is divergent when we extend to 
${\bf{g}}_2(F)$ respective $Gr_2(F_0)$ and the divergency is such that 
$c_1(X,Y)+(\delta b) (X,Y;F)$ will be finite. 

We will now try to find such a $b$. First we state:
\newtheorem{lemma1}{Lemma}
\begin{lemma1}
Let $b(X;F)=\frac{1}{8}\mbox{\rm {str}}\left( (F-F_0)F_0[F_0,X] _{(s)}\right)$ 
 for $X\in{\bf{g}}_1(F),F\in Gr_1(F_0)$. Then
\eq
\label{eq:OP}
(\delta b)(X,Y;F)=\frac{1}{8}\mbox{\rm str}\left( 
F\left[ \left[ F_0,X\right] _{(s)},\left[ F_0,Y\right] _{(s)} \right] _s\right).
\eqend
\end{lemma1}
{\bf Proof}
Using the definition of coboundary \Ref{eq:CB} we get 
\eqa
\label{eq:BR}
(\delta b)(X,Y;F) & = & \frac{1}{8}\mbox{str}\left( (F-F_0)F_0\left[ F_0,\left[ X,Y
\right] _s \right] _{(s)} \right)\nonu
&& - \frac{1}{8}\mbox{str}
\left(\left[ F,X\right] _{s} F_0\left[ F_0,Y\right] _{(s)} \right) \nonu
&&+\frac{1}{8}(-1)^{\mbox{\footnotesize{deg}}(X) \mbox{\footnotesize{deg}}(Y)}
\mbox{str}\left(\left[ F,Y\right] _{s} F_0\left[ F_0,X\right] _{(s)} \right).
\eqaend 
 Using the graded Jacobi identity for the first term while adding and 
subtracting the whole expression by the term 
\eq
  \frac{1}{8}\mbox{str}
\left(\left[ F_0,X\right] _{(s)} F_0\left[ F_0,Y\right] _{(s)} \right)
 - (-1)^{\mbox{\footnotesize{deg}}(X) 
\mbox{\footnotesize{deg}}(Y)}
\left( X\leftrightarrow Y\right)
\eqend
the right hand side of \Ref{eq:BR} can be written as
\eqa
\label{eq:EX}
&& \frac{1}{8}\mbox{str}\left( (F-F_0)F_0\left[ \left[ F_0,X\right] _{(s)},Y 
\right] _s \right) \nonu
&&+ \frac{1}{8}(-1)^{\mbox{\footnotesize{deg}}(X) 
\mbox{\footnotesize{deg}}(Y)}\mbox{str}
\left(\left[ F-F_0,Y\right] _{s} F_0\left[ F_0,X\right] _{(s)} \right)\nonu 
&&+\frac{1}{8}(-1)^{\mbox{\footnotesize{deg}}(X) \mbox{\footnotesize{deg}}(Y)}
\mbox{str}\left(\left[ F_0,Y\right] _{(s)} F_0\left[ F_0,X\right] _{(s)} 
\right) \nonu
&&- 
 (-1)^{\mbox{\footnotesize{deg}}(X) 
\mbox{\footnotesize{deg}}(Y)}
\left( X\leftrightarrow Y\right).
\eqaend
 With the formula for the supercommutator of a product: $[AB,C]_s=A[B,C]_s+
(-1)^{\mbox{\footnotesize{deg}}(B) \mbox{\footnotesize{deg}}(C)}[A,C]_sB$ 
we get
\eqa
&&(\delta b)(X,Y;F) = \frac{1}{8}\mbox{str}\left(\left[ (F-F_0)F_0
\left[ F_0,X\right] _{(s)},Y\right] _s \right)\nonu
&& -
 \frac{1}{8}(-1)^{\mbox{\footnotesize{deg}}(X) 
\mbox{\footnotesize{deg}}(Y)}\mbox{str}\left( (F-F_0)\left[ F_0,Y \right] 
_{(s)}\left[ F_0,X\right] _{(s)} \right) \nonu
&&+\frac{1}{8}(-1)^{\mbox{\footnotesize{deg}}(X) 
\mbox{\footnotesize{deg}}(Y)}\mbox{str}\left( \left[ F_0,Y \right] _{(s)}F_0
\left[ F_0,X\right] _{(s)} \right)\nonu
&& - (-1)^{\mbox{\footnotesize{deg}}(X) 
\mbox{\footnotesize{deg}}(Y)}\left( X\leftrightarrow Y\right).
\eqaend
 The first term vanish since it is the supertrace of a supercommutator. 
Since $F_0[F_0,Y]_{(s)}=-[F_0,Y]_{(s)}F_0$ the third term cancels the 
$F$-independent part of the second term. We are then left with  \Ref{eq:OP} 
which was to be proven.
\newtheorem{theorem2}{Theorem}
\begin{theorem2}
 For $X,Y\in{\bf{g}}_1(F)$, $F\in Gr_1(F_0)$ and $c_2(X,Y;F)=c_1(X,Y)+\delta b(X,Y;F)$ it holds that
\eq
\label{eq:MR}
c_2(X,Y;F)=
\frac{1}{8}\mbox{\rm str}_{\mbox{\rm c}} \left( (F-F_0)\left[\left[ F_0,X\right] 
_{(s)},\left[ F_0,Y\right] _{(s)}\right] _s\right)
\eqend
which exists even for $X,Y\in{\bf{g}}_2(F)$, $F\in Gr_2(F_0)$
\end{theorem2}
 The form of $c_2$ is obtained by adding the expressions \Ref{eq:LC} and 
\Ref{eq:OP}. That it is convergent for the given classes is easily seen by 
the following calculation:
\eqa
c_2(X,Y;F) & = & 
\frac{1}{16}\mbox{str} \left( \left((F-F_0) +F_0(F-F_0)F_0\right)
\left[\left[ F_0,X\right] _{(s)},
\left[ F_0,Y\right] _{(s)}\right] _s \right) \nonu
&& = -\frac{1}{16}\mbox{str} \left( (F-F_0)^2F_0
\left[\left[ F_0,X\right] _{(s)},
\left[ F_0,Y\right] _{(s)}\right] _s \right).
\eqaend  
 Thus it can be written as a supertrace of a trace class operator and is 
therefore well defined.
The function $c_2$ is a graded generalization of the Mickelsson-Rajeev cocycle 
\cite{MR}. 
 When $X\in{\bf{g}}_p(F),F\in Gr_p(F_0)$, $p>2$ the fundamental ideas are the 
same as for $p=2$, and for this reason we omit to study these cases. 
\section{A local form of the cocycles}
We will now determine the form of the cocycles 
for the case  $h=L^2({\bf R}^d)\otimes V$, where the finite dimensional 
${\bf Z}_2$-graded space $V$ 
introduces a natural grading in $h$ by: $f:{\bf R}^d\rightarrow 
V_{\alpha}\Rightarrow
 f\in h_{\alpha}$. Several of the calculations in this section are 
based on rules for PSDO's, summarized in the appendix. From now on 
we restrict to the case of $X$,$Y$ and $F$ being PSDO's of order $0$.
 The operators $X$ and $Y$ are also assumed to be 
compactly supported (in configuration space). We add 
the condition that the (super)commutator with $F_0$ for 
$X$ and $Y$ should to be a PSDO of 
order $-1$, while the corresponding condition for $F$ is that $F-F_0$ 
should be of order $-1$ (the motivation for these conditions will 
become clear later).
It is equivalent with the fact that the operators 
are in ${\bf{g}}_p(F_0)$ respective $Gr_p(F_0)$ for 
$2p=d+1$. This is a consequence of the fact that a PSDO 
on ${\bf R}^d$ is traceclass if and only if its symbol is of order $-d-1$ 
(see appendix). It means that when $d=1$ no regularization 
has to be done and the cocycle of interest is $c_1$. When $d=3$ the 
operators under consideration are in ${\bf{g}}_2$ and $Gr_2$, respectively. 
 A renormalization procedure has to be performed, leading to 
the cocycle $c_2$. 

 For a compactly supported PSDO $a$ of form \Ref{eq:AE} and of order $k$ 
the cut-off supertrace can be expanded as
\eqa 
\mbox{str}_\Lambda (a) & = & \int  d^dx\int\frac{d^dq}{(2\pi )^d} 
P_\Lambda (q)\mbox{str}_V\sigma (a) (q,x) \nonu
&&\!\!\!\!\!\!\!\!\!\!\!\!\!\!\!\!\!\!\!\!\!\!\!\!\!\!\! 
=  c_{k+d}(a)\Lambda ^{k+d} + ... +
 c_1(a)\Lambda + c_{\log }(a)\log (\frac{\Lambda}{\Lambda _0} ) \nonu
&&+ 
c_0(a)\Lambda ^0 +c_{-1}(a)\Lambda ^{-1} + ...
\eqaend
with $\Lambda _0$ some scale parameter, $P_\Lambda (q)=
\Theta (1-\frac{|q|}{\Lambda })$, and $\Theta $ is the Heaviside function.
The $\Lambda $-independent term $c_0(a)$ is of particular interest to us. 
We will refer to it as the regularized supertrace $\mbox{STR}(a)$. 
It is just the supertrace when the argument is a trace class operator. 
 The operators that will be considered here are such that $c_{\log }(a)$ is 
zero and thus the regularized supertrace is well defined.
 A PSDO that is compactly supported, homogeneous of order $-d$ and whose 
symbol can be written as a total derivative in momentum space has the 
property that 
the cut-off supertrace is $\Lambda $-independent and equal to the regularized 
supertrace of the operator. This is a nice property so we now try to 
find coboundaries that can be added to the cocycles 
\Ref {eq:LC} and \Ref {eq:MR} in a way that the resulting new cocycles can be 
written as a cut-off supertrace of such an operator.
 An example of a PSDO whose symbol to highest order can be written as a total 
derivative with respect to some momentum space variable is the 
supercommutator of two PSDO's
\eqa
\label{eq:HO}
&& \mbox{str}_V\left(\sigma\left(\left[ a,b\right] _s\right)\right) (q,x)\nonu
&= & -i\sum_{k=1}^d\mbox{str}_V\Bigg(\frac{\partial}{\partial q_k}\left(
\sigma (a)  \frac{\partial}{\partial x_k}\sigma (b)\right) \nonu
&&-
(-1)^{\mbox{\footnotesize{deg}}(a) \mbox{\footnotesize{deg}}(b)}
\frac{\partial}{\partial x_k}\left(\left(\frac{\partial}{\partial q_k}
\sigma (b) \right) \sigma (a) \right)\Bigg)(q,x)\nonu 
&&+\mbox{lower order terms.}
\eqaend
 This is shown by using the formula \Ref{eq:PR} for calculating the symbol of 
the product of two PSDO's. The second term in the expression is uninteresting to
us since it vanishes when integrating if one of the operators under 
consideration is compactly supported. The lower order terms in the expression 
can also be written as a sum of total derivatives of momentum and 
configuration variables. 
Motivated by the discussion above we try to find some functions 
$b_1$ and $b_2$ such that the new cocycles 
\eqa
\label{eq:2C}
\tilde{c}_1 & = & c_1- \delta b_1 \nonu
\tilde{c}_2 & = & c_2-\delta b_2-\delta b_1
\eqaend 
can be written as a regularized supertrace of a supercommutator. 

The choice 
\eqa
b_1 & = & \frac{1}{2}\mbox{STR}\left( F_0X \right) \nonu
b_2 & = & \frac{1}{8}\mbox{STR}\left( FF_0[F_0,X]_{(s)} \right)
\eqaend  
will do if the restriction 
$\sigma(F_0)(q,x)=\sigma(F_0)(q)$ is put on $F_0$. From a physical point 
of view this is reasonable if thinking of $F_0$ as the sign of the free 
Hamiltonian. 
It implies that
\eq
\mbox{STR}(F_0aF_0)=\mbox{STR}(a)
\eqend
is true for $a$ trace class. 
The relation 
\eq
\label{eq:SS}
\mbox{str}_{\mbox{c}}(a)=\mbox{STR}(a) 
\eqend
holds therefore if $a$ is conditional trace class 
since it obviously holds when  $a$ is trace class. 
Now, using \Ref {eq:LC}, \Ref {eq:HO}--\Ref {eq:SS} and the linearity of the 
regularized supertrace, we obtain
 \eqa
\label{eq:C1}
&&\tilde{c}_1(X,Y) = -\frac{1}{2}\mbox{STR}
\left( 
\left[ X,F_0Y\right] _s\right) \nonu
& = & 
\frac{i}{2}\int dx\int _{|q|\leq\Lambda }\frac{dq}{2\pi } 
\mbox{str}_V \frac{\partial}{\partial q}\Big(
\sigma (X)_0(q,x)\nonu
&&\frac{\partial}{\partial x}\Big(\sigma \left( F_0\right) _0
\left( q\right) 
\sigma \left( Y\right) _0 \left( q,x\right)\Big) \Big) 
\eqaend
where $\sigma (\cdot )_k$ stands for the piece of the symbol that is 
homogeneous of order $k$.
 Similarly, using 
\eqa
&&\mbox{str}_{\Lambda}\left(\left[ \left[ F_0,X\right] _{(s)},
Y\right]_s \right)- (-1)^{\mbox{\footnotesize{deg}}(X) 
\mbox{\footnotesize{deg}}(Y)}
\left( X\leftrightarrow Y\right)\nonu 
& = &
\mbox{str}_{\Lambda}\left(\left[ F_0,\left[ X,Y\right] _s\right] _{(s)}
\right)=0
\eqaend
and the $F$-dependent part of the calculation in the proof of Lemma 1 with 
the supertrace replaced by the regularized supertrace (remember that the 
regularized supertrace of a supercommutator is not necessary zero and the 
expression in \Ref{eq:EX} is therefore useful) together 
with \Ref {eq:MR} and \Ref {eq:2C}--\Ref {eq:SS} gives
\eqa
\label{eq:C2}
&&\tilde{c}_2(X,Y;F)  =  \frac{1}{8}\mbox{STR}\left( (F-F_0)
\left[ F_0,X\right] _{(s)}\left[ F_0,Y\right] _{(s)}\right)
\nonu
&&-\frac{1}{8}\mbox{STR}\left( FF_0\left[\left[ F_0,X\right] _{(s)},
Y\right]_s \right) + \frac{1}{8}\mbox{STR}\left(
\left[ F,X\right] _s F_0\left[ F_0,Y\right] _{(s)} \right)\nonu
&&-
 \frac{1}{2}\mbox{STR}\left( F_0XY\right)-
(-1)^{\mbox{\footnotesize{deg}}(X) \mbox{\footnotesize{deg}}(Y)}
\left( X\leftrightarrow Y\right) \nonu
& = & -\frac{1}{4}\mbox{STR}\left( \left[ X,F_0Y\right] _s\right)
-\frac{1}{8} \mbox{STR}\left(\left[ (F-F_0)F_0\left[ F_0,X\right] 
_{(s)},
Y\right]_s \right)\nonu
&& - (-1)^{\mbox{\footnotesize{deg}}(X) 
\mbox{\footnotesize{deg}}(Y)}
\left( X\leftrightarrow Y\right). 
\eqaend
 Assuming that the first term vanishes (as it will in the special cases 
below) we use \Ref {eq:HO} to obtain 
\eqa
\label{eq:CF}
&&\tilde{c}_2(X,Y;F) \nonu 
&&=  \frac{i}{8}\sum_{k=1}^3\int d^3x\int _{|q|\leq\Lambda }\frac{d^3q}
{(2\pi )^3 }\mbox{str}_V
\frac{\partial}{\partial q_k}\nonu
&&\left(  
 \sigma (F-F_0)_{-1}(q,x) \sigma(F_0)_0(q)
\sigma\left(\left[ F_0,X\right] _{(s)}\right)_{-1}(q,x) 
\frac{\partial}{\partial x_k}\sigma \left( Y\right)_0 (q,x)\right) \nonu 
&&- (-1)^{\mbox{\footnotesize{deg}}(X) 
\mbox{\footnotesize{deg}}(Y)}
\left( X\leftrightarrow Y\right).
\eqaend 
\section{Two examples}
 Some concrete examples will now be considered. The operators of interest are 
the Hamiltonian $H=H_0+A$ and compactly supported gauge transformations 
$e^{X}$. The free Hamiltonian $H_0$, $A$ and 
$X$ are assumed to be PSDO's such that $\sigma ({H_0}^2)(q,x)=|q|^2$, while 
$A$ and $X$ are assumed to be of order $0$. 
 Let $F_{(0)}=\mbox{sgn}(H_{(0)})$, using the spectral theorem with 
$\mbox{sgn}(x)=1$ $(-1)$ for $x\geq 0$ $(x<0)$.
 Then 
\eqa
&&\sigma (H^2)(q,x)\nonu
&&=|q|^2\left( 1+\frac{\sigma (H_0)\sigma (A)+
\sigma (A)\sigma (H_0)}{|q|^2}\right)(q,x)+{\cal O}(|q|^0)
\eqaend
and using \Ref{eq:PR},
\eqa
&&\sigma (1/\sqrt{H^2})(q,x)\nonu
&&=\frac{1}{|q|}\left( 1-\frac{\sigma (H_0)
\sigma (A)+
\sigma (A)\sigma (H_0)}{2|q|^2}\right)(q,x)+{\cal O}(|q|^{-3})
\eqaend
implying 
\eqa
\label{eq:FF}
&&\sigma (F)(q,x)=\nonu
&&\frac{1}{|q|}\left(\sigma (H_0)+\frac{\sigma (A)}{2}
-\frac{\sigma (H_0)\sigma (A)\sigma (H_0)}{2|q|^2}\right)(q,x)+
{\cal O}(|q|^{-2}).
\eqaend
Thus we get that $F-F_0$ is a PSDO of order $-1$. 
 
 Performing an infinitesimal gauge transformation gives 
\eq
\sigma(H_0+A)\rightarrow \sigma(H_0+A_X) =
 \sigma(H_0+A-[A,X]_s-[H_0,X]_{(s)})+{\cal O}(|q|^{-1}).
\eqend
 Since both $A$ and $A_X$ are PSDO's of order 0 we see that a gauge 
transformation must obey the condition that $[H_0,X]_{(s)}$ is of order 0. 
 This in turn implies that $[F_0,X]_{(s)}$ is of order $-1$. 

 The way the operators have been introduced here they fulfill all 
restrictions put in the former section. The results obtained there can  
therefore be used. Two different choices of 
$H_0$, $A$ and $X$ will be considered in the next section. For each 
choice the cocycles \Ref {eq:C1} and \Ref {eq:CF} will be calculated. 
 Note that we only have to specify $H_0$, $A$ and $X$ to highest order 
(as PSDO's) since it is only this part of them that appears in the formulas 
for the cocycles. 
 In both cases $V=V_{\mbox{spin}}\otimes V_{\mbox{color}}\otimes 
V_{{\bf Z}_2}$, that is, 
the boson and fermion wavefunctions will be assumed to have an equal number of 
spin and color (internal degrees) components. It is useful to define 
some additional (super)traces: $\mbox{str}_V=\mbox{tr}_{\mbox{spin}}
\mbox{str}_{\mbox{color}}=
\mbox{tr}_{\mbox{spin}}\mbox{tr}_{\mbox{color}}\mbox{str}_{{\bf Z}_2}$. 

The equations \Ref {eq:C1} and \Ref {eq:CF} for the cocycles has a momentum 
integration over a domain that contains the origin. Now, since there are 
 factors $\frac{1}{|q|}$ in the integrand, coming from $\sigma (F_{(0)})$,
 the integral is not well defined. To avoid this problem the factors 
 $\frac{1}{|q|}$ will be regularize close to the origin in momentum space.
 It turns out (see below) that the formulas for the cocycles are independent 
of how such an infrared regularization is made. It is so since Stokes' 
theorem may be used and thus the cocycles depend only on the 
behavior of the integrand in a neighborhood of the boundary of the 
integration domain in momentum space. We denote our regularized function 
by $\frac{1}{|q|}_r$ and define it to be equal to $\frac{1}{|q|}$ when 
$|q|\geq \delta$ and some smooth function when $|q|\leq \delta$.
\subsection{Case 1}
 The operators will here be written in a matrix form according to the 
decomposition $h=h_{\bar{0}}\oplus h_{\bar{1}}$, just as was done in 
\Ref {eq:MX}. To get a rather simple form of the cocycles \Ref {eq:C1} and 
\Ref {eq:CF}  we mimic the chiral fermion case and 
assume that also the energy spectrum in the boson space is determined by the 
fermion Hamiltonian. We choose to consider vector potentials that are of the 
same form in all four sectors and it should be a generalization of the 
\lq ordinary\rq \, potential in the fermion to fermion sector. Thus:
\[
\sigma (H_0)(q,x)  = 
\left(
 \begin{array}{cc}
 q\!\!\! / &  0 \nonu
  0 &   q\!\!\! /
 \end{array} \right)
\quad\sigma (A)(q,x)= A\!\!\! / (x)=
\left(
 \begin{array}{cc}
 {A \!\!\! /}_{\bar{0}\bar{0}}(x)  &  {A \!\!\! /}_{\bar{0}\bar{1}}(x) \nonu
  {A \!\!\! /}_{\bar{1}\bar{0}}(x) &  {A \!\!\! /}_{\bar{1}\bar{1}}(x) 
 \end{array} \right) 
\]
\eq
\label{eq:1H}
\mbox{where}\quad S\!\!\!\! /   = \left\{ \begin{array}{ll}
S & d=1 \\
\sum_{i=1}^3S_j\sigma _j & d=3 
 \end{array}\right.
\quad S=q,A,A_{\alpha\beta }(x), \quad \alpha ,\beta\in {\bf Z}_2
\eqend 
where $\sigma _j$, $j=1,2,3$ are the usual Pauli spin matrices 
and the $A_{\alpha\beta }$'s are assumed to commute with these.
 Motivated by the 
ungraded case (the \lq usual\rq \, gauge transformations leading to the 
conservation of the electric charge) we restrict ourselves to the case of $X$ 
being a multiplication operator; $(Xf)(x)=X(x)f(x)$, commuting with the Pauli 
matrices.

When $d=1$ 
 a simple calculation leads to
\eqa
\tilde{c}_1(X,Y) & = & \frac{i}{2}\int  dx\int _{|q|\leq\Lambda }
\frac{dq}{2\pi } \mbox{str}_V \frac{\partial}{\partial q}\left(
X(x)\frac{\partial}{\partial x}\left(\frac{ q\!\!\! /}{|q|_r}Y(x)\right) 
\right) 
\nonu 
&&=\frac{i}{2\pi }\int  dx\mbox{str}_{\mbox{color}}\left( \left( 
\frac{\partial}{\partial x}X(x) \right) Y(x)\right).
\eqaend
 This is recognized as a graded generalization of the affine Kac-Moody cocycle 
(see ref.~\cite{KR}).

Let us now consider the case $d=3$. Since $\mbox{tr} (\sigma _j)=0$, 
$j=1,2,3$, the first term in \Ref {eq:C2} vanishes and thus we may use 
equation \Ref{eq:CF}.
Since 
\eq
\sigma \left( \left[ F_0, X \right]_{(s)} \right) (q,x) = \sum_{j=1}^3
 \left( -i \frac{\partial}{\partial q_j}\frac{q \!\!\! /}{|q|_r}\right) 
\frac{\partial}{\partial x_j}X(x)+{\cal O}(|q|^{-2})
\eqend
and
\eq
\sigma \left( F-F_0 \right)(q)= \frac{1}{|q|_r}\sum_{i=1}^3A_i(x)
\left( \sigma _i-\frac{q  \!\!\! / q_i}{|q|_r^2}\right)+{\cal O}(|q|^{-2})
\eqend
we obtain 
\eqa
\label{eq:CJ}
\tilde{c}_2(X,Y;A) & = & -\frac{1}{8}\sum_{i,j,k=1}^3\mbox{str}_
{\mbox{color}}
J_{ijk}^{\Lambda}\int  d^3x\left( A_i(x) \frac{\partial X}
{\partial x_j}(x)\frac{\partial Y}{\partial x_k}(x)\right)\nonu
&&- (-1)^{\mbox{\footnotesize{deg}}(X) 
\mbox{\footnotesize{deg}}(Y)}
\left( X\leftrightarrow Y\right)
\eqaend
where 
\eq
J_{ijk}^{\Lambda}=-\int _{|q|\leq\Lambda }\frac{d^3q}{(2\pi )^3 }
\mbox{tr}_{\mbox{spin}} \frac{\partial}{\partial q_k}\left( 
\left(\sigma _i-\frac{q  \!\!\! / q_i}{|q|_r^2}\right)
\frac{q \!\!\! /}{|q|_r} \left(
\frac{\partial }{\partial q_j}\frac{q \!\!\! /}{|q|_r}\right)\right).
\eqend
 Since $J_{ijk}^{\Lambda}$ 
is of the form of an integration over a derivation we may replace the 
expression $\frac{\partial |q|_r}{\partial|q|}$ with $1$ whenever it occurs 
in the integrand as long as $\Lambda$ is greater than the small 
regularization parameter $\delta$. Therefore the relation 
$\frac{\partial}{\partial q_i}\frac{q  \!\!\! /}{|q|_r}=
\frac{1}{|q|}\left( \sigma _i-\frac{q\!\!\! /q_i}{|q|_r^2}\right)$ holds 
under the integration. 
 Using this together with $\mbox{tr}_{\mbox{spin}}(\sigma _i\sigma _j
\sigma _k)=2i\epsilon _{ijk}$ and Stokes' theorem gives
\eq
\label{eq:JE}
J_{ijk}^{\Lambda}=-\int _{|q|\leq\Lambda }\frac{d^3q}{(2\pi )^3 }
\frac{\partial}{\partial q_k}\left( 2i\epsilon_{ilj}\frac{q_l}
{|q|^3}\right)=-2i\epsilon _{ilj}\frac{\delta _{kl}}{6\pi ^2}=
\frac{i}{3\pi ^2}\epsilon _{ijk}
\eqend
which combined with \Ref {eq:CJ} results in
\eqa
\label{eq:2F}
\tilde{c}_2 (X,Y;A) & = & -\frac{i}{24\pi ^2} \sum_{i,j,k=1}^3\int  d^3x
\mbox{str}_{\mbox{color}} \left(\epsilon _{ijk} A_i(x) \frac{\partial X}
{\partial x_j}(x)\frac{\partial Y}{\partial x_k}(x) \right)\nonu
&&- (-1)^{\mbox{\footnotesize{deg}}(X) 
\mbox{\footnotesize{deg}}(Y)}
\left( X\leftrightarrow Y\right).
\eqaend 
 This is recognized as a graded generalization of the Mickelsson-Faddeev
 cocycle\cite{M,F}. The calculation 
performed above is close to a corresponding calculation\cite{LM} of the 
Schwinger term for chiral fermions.
\subsection{Case 2}
Massless bosons can be described by a vector-valued (needed for the spin and 
the internal degrees of freedom) function $\varphi$ 
obeying the Klein-Gordon equation 
\eq 
\Box \varphi =0.
\eqend 
Introduce the conjugate momentum, $\pi ^{\ast}=\frac{\partial}{\partial t}
\varphi $, in order to get an equivalent description containing two-component 
functions with the Hamiltonian 
 \eq
\tilde{H}_0=
\left(
 \begin{array}{cc}
 0 & i  \nonu
 -i|D|^2 & 0 
 \end{array} \right)
\eqend
with $|D|^2=-\triangle$, the Laplacian, so that:
\eq
i\frac{\partial}{\partial t}\left(
\begin{array}{c}
\varphi \nonu
\pi ^{\ast}
\end{array} \right)=\tilde{H}_0
\left(
\begin{array}{c}
\varphi \nonu
\pi ^{\ast}
\end{array} \right).
\eqend  
 Combining bosons with chiral fermions leads to the study of the Hamiltonian 
\eq
\label{eq:2H}
H_0=
\left(
 \begin{array}{ccc}
 0 & i & 0 \nonu
 -i|D|^2 & 0 & 0\nonu
0 & 0 & D 
 \end{array} \right)
\eqend
with $\sigma (D) (q,x) = q  \!\!\! /$, with $q  \!\!\! /$ as in the former 
example,  acting on column vectors of form
\eq
\label{eq:CV}
\left(
\begin{array}{c}
\varphi \nonu
\pi ^{\ast} \nonu
\psi
\end{array} \right)
\eqend
where $ \psi$ denotes the fermionic wavefunction. 
The case when the fields $\varphi$, 
$\pi ^{\ast}$ and $\psi$ have an equal number of spin components will be 
considered. 
We choose to consider gauge transformations $X$ of the form
\eq
X(x) =\left(
\begin{array}{ccc}
X_{\varphi\varphi }(x) & 
i X_{\varphi\pi ^{\ast} }(x)\frac{D}{ |D|^2}
& X_{\varphi\psi }(x) \nonu
-iD X_{\varphi\pi ^{\ast} }(x) & 
DX_{\varphi\varphi }(x)\frac{D}{ |D|^2} & 
-iDX_{\varphi\psi }(x)  \nonu
X_{\psi\varphi }(x)
& iX_{\psi\varphi }(x)\frac{D}{ |D|^2} & X_{\psi\psi }(x)
\end{array} \right)
\eqend 
where the $X_{\phi _{1}\phi _{2}}(x)$'s are multiplication operators 
commuting with the Pauli matrices. 
The motivation for this choice is the following. We would like to have 
multiplication operators commuting with the Pauli matrices that mixes the 
$\varphi$ and $\psi$ among themselves. 
This motivates the choice of the four elements in the corners of the matrix. 
 The form of the remaining elements (to highest order) follow directly from 
this and the condition  that the (super)commutator with $H_0$ should be of 
order $0$ together with the claim that the form of the gauge transformation 
should be preserved under the supercommutator. 
 Taking the (super)commutator with $H_0$ gives
 \eq
[H_0,X(x)]_{(s)} =\left(
\begin{array}{ccc}
[D,X_{\varphi\pi ^{\ast}}(x)] &
i [D,X_{\varphi\varphi }(x)]\frac{D}{ |D|^2}
& [D,X_{\varphi\psi }(x)] \nonu
-iD [D , X_{\varphi\varphi }(x)] & 
D [ D , X_{\varphi\pi ^{\ast} } (x) ] \frac{D}{ |D|^2} & 
-iD [ D , X_{\varphi\psi } (x) ]  \nonu
 \left[ D , X_{\psi\varphi } (x) \right]
& i[D,X_{\psi\varphi }(x)]\frac{D}{ |D|^2} & [D,X_{\psi\psi }(x)]
\end{array} \right)
\eqend 
which motivates us to consider vector potentials of form 
\eq
A \!\!\! /(x) =\left(
\begin{array}{ccc}
A \!\!\! / _{\varphi\varphi }(x) & 
i A \!\!\! /_{\varphi\pi ^{\ast} }(x)\frac{D}{ |D|^2}
& A \!\!\! /_{\varphi\psi }(x) \nonu
-iD A \!\!\! /_{\varphi\pi ^{\ast} }(x) & 
DA \!\!\! /_{\varphi\varphi }(x)\frac{D}{ |D|^2} & 
-iDA \!\!\! /_{\varphi\psi } (x) \nonu
A \!\!\! /_{\psi\varphi }(x)
& iA \!\!\! /_{\psi\varphi }(x)\frac{D}{ |D|^2} & A \!\!\! /_{\psi\psi }(x)
\end{array} \right)
\eqend
where the $A \!\!\! /_{\phi _{1}\phi _{2}}(x)$'s are as in the former section 
(in the case $d=3$ they are a sum of Pauli matrices multiplied with 
multiplication operators that commutes with the Pauli matrices). 

To easier see the structure of the operators, they will be conjugated by
\eq
W=\frac{1}{\sqrt{2}}\left(
\begin{array}{ccc}
|D|^{-1/2} & |D|^{-1/2} & 0 \nonu
-i\frac{D}{|D|^{1/2}} & i\frac{D}{|D|^{1/2}} & 0 \nonu
0 & 0 & \sqrt{2}
\end{array} \right)
\eqend
according to 
\eqa
\Phi & \rightarrow & \Phi ^W=W^{-1}\Phi \nonu
H_0 & \rightarrow & H_0^W =W^{-1}H_0W\nonu
X & \rightarrow & X^W =W^{-1}XW\nonu
A & \rightarrow & A^W =W^{-1}AW.
\eqaend
 Then we get
\eq
\label{eq:HW}
H_0^W =\left(
\begin{array}{ccc}
D & 0 & 0 \nonu
0 & -D& 0 \nonu
0 & 0 & D
\end{array} \right)
\eqend
\eqa
\label{eq:XW}
X^W(x) & = & \left( 
\begin{array}{c}
|D|^{1/2} \left( X_{\varphi\varphi }(x) +  X_{\varphi\pi ^{\ast} }(x) \right)
 |D|^{-1/2}   \nonu
0  \nonu
\sqrt{2}X_{\psi\varphi }(x) |D|^{-1/2}  \nonumber 
\end{array} \right. \nonu
&&\left.\begin{array}{cc}  
0 &\sqrt{2} |D|^{1/2} X_{\varphi\psi }(x)  \nonu
|D|^{1/2} \left( X_{\varphi\varphi }(x)  -X_{\varphi\pi ^{\ast} }(x) \right) 
 |D|^{-1/2}  & 0 \nonu
0 & X_{\psi\psi }(x) 
\end{array} \right) 
\eqaend
\eqa
A \!\!\! /^W(x)& = &\left( 
\begin{array}{c}
|D|^{1/2} \left( A \!\!\! /_{\varphi\varphi }(x) + 
 A \!\!\! /_{\varphi\pi ^{\ast} }(x)\right) |D|^{-1/2}   
 \nonu
0 \nonu
\sqrt{2}A \!\!\! /_{\psi\varphi }(x)|D|^{-1/2}  
\end{array} \right.\nonu
&&
\left. 
\begin{array}{cc}
0 & \sqrt{2}|D|^{1/2} A \!\!\! /_{\varphi\psi }(x) \nonu
|D|^{1/2} \left( A \!\!\! /_{\varphi\varphi }(x) 
-A \!\!\! /_{\varphi\pi ^{\ast} }(x)\right) |D|^{-1/2}  & 0 \nonu
0 & A \!\!\! /_{\psi\psi } (x)
\end{array} \right. .
\eqaend 
 It is easy to check that the formulas for the cocycles \Ref {eq:C1} and 
\Ref {eq:CF}  are independent of such a conjugation (use \Ref {eq:FF} and 
identities as $|H^W_0|=|H_0|$). Thus, $W$-conjugated operators can be used 
in the calculations. 

For $d=1$ we use \Ref {eq:HW} and \Ref {eq:XW} in \Ref {eq:C1} to get 
\eq
\tilde{c}_1(X,Y) =\frac{i}{2}\int  dx\int _{|q|\leq\Lambda }\frac{dq}{2\pi } 
\mbox{str}_V \frac{\partial}{\partial q}T(q,x)
\eqend
where $T$ is a matrix $\{ T_{ij} \} _{ij}$ with the diagonal entries 
\eqa
T_{11}(q,x) & = & |q|_r^{-1/2}\left( X_{\varphi\varphi }
+  X_{\varphi\pi ^{\ast} }\right) \frac{q}{|q|_r} \frac{\partial}{\partial x}
\left( Y_{\varphi\varphi }+  Y_{\varphi\pi ^{\ast} }\right) |q|^{1/2} \nonu
&&+2|q|^{1/2} X_{\varphi\psi }\frac{q}{|q|_r} \frac{\partial}{\partial x}
Y_{\psi\varphi }|q|_r^{-1/2} \nonu
T_{22}(q,x) & = & -|q|_r^{-1/2}\left( X_{\varphi\varphi }
-  X_{\varphi\pi ^{\ast} }\right) \frac{q}{|q|_r} \frac{\partial}{\partial x}
\left( Y_{\varphi\varphi }-  Y_{\varphi\pi ^{\ast} }\right) |q|^{1/2} \nonu
T_{33}(q,x) & = & 2X_{\psi\varphi }\frac{q}{|q|_r} \frac{\partial}{\partial x} 
Y_{\varphi\psi } 
+X_{\psi\psi } \frac{q}{|q|_r}\frac{\partial}{\partial x} Y_{\psi\psi }.
\eqaend
 The factor $\frac{\partial}{\partial q}\frac{q}{|q|_r}$ might be replaced  
by $2\delta (q)$ under the momentum integration leading to a 
 trivial $q$-integration. Evaluating the supertrace and collecting 
terms gives 
\eqa
\tilde{c}_1(X,Y) & = & \frac{i}{2\pi }\int  dx \mbox{tr}_{\mbox{color}} \Big( 
2X_{\varphi\pi ^{\ast} }\frac{\partial }{\partial x}Y_{\varphi\varphi } 
+2X_{\varphi\varphi } \frac{\partial}{\partial x}Y_{\varphi\pi ^{\ast} } \nonu
&& +  2X_{\varphi\psi } \frac{\partial }{\partial x}Y_{\psi\varphi }
-2X_{\psi\varphi }\frac{\partial}{\partial x}Y_{\varphi\psi } 
-X_{\psi\psi }\frac{\partial}{\partial x} Y_{\psi\psi } \Big)
\eqaend
where $\mbox{tr}_{\mbox{color}}$ is defined by  $\mbox{tr}_{\mbox{color}}
\mbox{str}_{{\bf{Z}}_2}=\mbox{str}_{\mbox{color}}$, 
and $\mbox{str}_{{\bf{Z}}_2}$ is defined on matrices 
$\{ a_{ij} \} _{ij}$ of form \Ref {eq:XW} by 
\eq
\mbox{str}_{{\bf{Z}}_2}(a)=a_{11}+a_{22}-a_{33}.
\eqend
When $d=3$, the first term in \Ref {eq:C2} 
will vanish by the same reason as for the former case .

 We now get
\eqa
&&\sigma \left( \left[ F_0^W, X^W \right]_{(s)} \right) (q,x) \nonu & = &  
\sum_{j=1}^3\left(-i\frac{\partial}{\partial q_j}
 \sigma (F_0^W)(q)\right) \frac{\partial}{\partial x_j}
\sigma (X^W)_0(q,x)+{\cal O}(|q|^{-2})
\eqaend
and
\eq
\sigma \left( F^W-F_0^W \right)(q)= 
\frac{1}{|q|_r}\sum_{i=1}^3\sigma (A_i^W)_0(q,x)
\left( \sigma _i-\frac{q  \!\!\! / q_i}{|q|_r^2}\right)
+{\cal O}(|q|^{-2})
\eqend
 leading to
\eqa
\label{eq:CE}
\tilde{c}_2(X,Y;A) & = & -\frac{1}{8}\sum_{i,j,k=1}^3
J_{ijk}^{\Lambda}\int  d^3x\nonu
&&\mbox{str}_{\mbox{color}}\left( \sigma (A_i^W)_0
(q,x) \frac{\partial \sigma (X^W)_0}
{\partial x_j}(q,x)\frac{\partial \sigma (Y^W)_0}{\partial x_k}(q,x)\right)
\nonu
&&- (-1)^{\mbox{\footnotesize{deg}}(X) 
\mbox{\footnotesize{deg}}(Y)}
\left( X\leftrightarrow Y\right)
\eqaend 
where the expression under the $x$-integration actually is $q$-independent. 
Introducing the operators 
\eqa
\tilde{X}(x) & = & \left(
\begin{array}{ccc}
X_{\varphi\varphi }(x)+  X_{\varphi\pi ^{\ast} }(x)  & 0 
&\sqrt{2} X_{\varphi\psi }(x) 
\nonu
0 &  X_{\varphi\varphi }(x) -X_{\varphi\pi ^{\ast} }(x) & 0 \nonu
\sqrt{2}X_{\psi\varphi }(x) & 0 & X_{\psi\psi }(x)
\end{array} \right) \nonu
\tilde{A}\!\!\! /(x) & = & \left( 
\begin{array}{ccc}
 A \!\!\! /_{\varphi\varphi }(x) +  A \!\!\! /_{\varphi\pi ^{\ast} }(x)
 & 0 & \sqrt{2}A \!\!\! /_{\varphi\psi }(x) \nonu
0 & \ A \!\!\! /_{\varphi\varphi }(x) -A \!\!\! /_{\varphi\pi ^{\ast} }(x)  
& 0 \nonu
\sqrt{2}A \!\!\! /_{\psi\varphi }(x)  & 0 & A \!\!\! /_{\psi\psi } (x)
\end{array} \right)
\eqaend
and using \Ref {eq:JE} the equation above can be rewritten as
\eqa
\tilde{c}_2 (X,Y;A) & = & -\frac{i}{24\pi ^2} \sum_{i,j,k=1}^3\int  d^3x
\mbox{str}_{\mbox{color}} \left(\epsilon _{ijk} \tilde{A}_i(x) 
\frac{\partial \tilde{X}}
{\partial x_j}(x)\frac{\partial \tilde{Y}}{\partial x_k}(x) \right) \nonu
&&- (-1)^{\mbox{\footnotesize{deg}}(X) 
\mbox{\footnotesize{deg}}(Y)}
\left( X\leftrightarrow Y\right)
\eqaend 
which is similar to the result  \Ref {eq:2F} in the former example.

\thanks{\bf Acknowledgments:}


I am grateful to J. Mickelsson for interesting me in the field under 
consideration.
I would also like to thank him and E. Langmann for many useful discussions. 
\appendix
 \section*{Pseudodifferential operators (PSDO)}
\label{sec:PSDO}
Here some facts about PSDO will be summarized (see 
ref.~\cite{H} and  ref.~\cite{T}). A PSDO $a$ on the Hilbert 
space $ L^2({\bf R}^d)\otimes V$, $V$ a finite dimensional vector space and 
$N=\mbox{dim}(V)$, is 
determined by its symbol $\sigma (a)$ which is a smooth $N \times N$ matrix  
valued function, such that
\begin{equation}
\left(af\right)(x)=\frac{1}{\left(2\pi\right)^{d/2}}\int d^dq\sigma (a)
(q,x) \hat{f}(q)e^{iq\cdot x} 
\end{equation}
where   
\eq
\hat{f}(q)=\frac{1}{\left(2\pi\right)^{d/2}}\int d^dxf (x)e^{-iq\cdot x} 
\quad  f\in L^2({\bf R}^d)\otimes V.
\eqend

The PSDO we are concerned with admits asymptotic expansion of their symbols as
\begin{equation}
\label{eq:AE}
\sigma (a)(q,x) \sim \sigma (a)_k(q,x)+ \sigma (a)_{k-1}(q,x)+ 
\sigma (a)_{k-2}(q,x)+...
\end{equation}
where $k$ is an integer and 
\eq
\sigma (a)_j(\mu q,x)=\mu ^j\sigma (a)_j(q,x), \quad \mbox{for }
\mu>1, |q|\gg 1.
\eqend
 Such a PSDO is said to be of order $k$. A PSDO $a$ that has only one term in 
its expansion
\eq
\sigma (a)(q,x) \sim \sigma (a)_k(q,x)
\eqend
is said to be homogeneous of order $k$.

The symbol for the product of two PSDO's is given asymptotically by 
\begin{equation}
\label{eq:PR}
\sigma (ab) (q,x) \sim \sum \frac{(-i)^{|m|}}{m!}\left(\partial_q^m 
\sigma (a)(q,x) 
\partial_x^m \sigma (b)(q,x)\right) 
\end{equation}
where the sum is over all sets of nonnegative integers $m=(m_1,...,m_d), 
\quad |m|=m_1+...+m_d, \quad \partial_x^m= 
\frac{\partial ^{m_1}}{\partial x_1} ... \frac{\partial ^{m_d}}{\partial x_d}  
$, etc., and $m!=m_1!...m_d!$.

Finally, the trace of a trace class operator in ${\bf R}^d$ is given by
\begin{equation}
\mbox{tr}a=\frac{1}{(2\pi)^d} \int d^dxd^dq \mbox{tr}_V \sigma (a)(q,x).
\end{equation}
 Thus, an operator with compact support in configuration space 
is Hilbert-Schmidt if and only if it has a symbol of order $-d/2$ or less.

\end{document}